\documentclass[prb,twocolumn,showpacs,preprintnumbers,amsmath,amssymb,superscriptaddress]{revtex4-1}

\newcommand{\tht}{\theta}
\newcommand{\aaa}{\alpha}
\newcommand{\bbb}{\beta}
\newcommand{\no}{\nonumber}
\usepackage{bm}
\usepackage{url} 
\usepackage{color}
\usepackage{amsmath,graphicx}

\begin{document}
\title{Simple approximate formulas for postbuckling deflection of heavy elastic columns}

\author{Hiroyuki Shima}
\email{hshima@yamanashi.ac.jp}
\thanks{(Correspondence author)} 
\affiliation{Department of Environmental Sciences, University of Yamanashi, 4-4-37, Takeda, Kofu, Yamanashi 400-8510, Japan}

\date{\today}

\begin{abstract}
Columnar buckling is a ubiquitous phenomenon that occurs in both living things and man-made objects, regardless of the length scale ranging from macroscopic to nanometric structures. In general, analyzing the post-buckling behavior of a column requires the application of complex mathematical methods because it involves nonlinear problem solving. To complement these complex methods, this study presents simple analytical formulas for the large deflection of a heavy elastic column under combined loads. The analytical formulas relate the concentrated load acting on the tip of the column, the column's own weight, and the deflection angle of the column through a simple mathematical expression. This can assist in obtaining an overall picture of the post-buckling behavior of heavy columns from an application point of view.
\end{abstract}

\maketitle
\section{Introduction}

When a vertical slender column is axially compressed, it may suddenly flex archwise and lose its ability to withstand the load. This deflection, called columnar buckling, is a crucial consideration in structural design because the column becomes unstable under compressive stress considerably lesser than the failure strength of the constituent materials \citep{Brush1975}. In addition to the field of structural design, columnar buckling is relevant in a wide range of fields in which slender column structures are involved; plant science, nanomaterial engineering \citep{Chinnawut2020}, and robotics \citep{Sipos2020} are only a few to mention. In plant science, for example, columnar buckling has been presumed to impose an upper limit on the growth height of upright stems and trunks \citep{McMahon1973,Niklas1994,Dargahi2019}. Plants are subjected to self-weight-induced downward load and concentrated loads caused by {\it e.g.,} the petals, leaves, or rain water accumulated near the tip; therefore, plant growth is considered to be regulated for avoiding buckling under these combined loads. Further, it has been suggested that columnar buckling occurs in nanometric tubes and wires \citep{GWWang2004,ShimaMaterial2012,Mustapha2012,JXWu2015,Robinson2019,Umeno2019}, whose mechanical stability is crucial for improving the performance of nanoelectromechanical devices \citep{Carr2003,Roodenburg2009,Weick2011}.

From a mathematical point of view, columnar buckling is the bifurcation of the solution of a nonlinear differential equation. This nonlinear equation governs the static equilibrium states of an elastic column under axial compression. Bifurcation signifies the transition of the equilibrium column shape from a vertically standing state to a deflected (postbuckled) one. This transition occurs at critical values of the variable coefficients involved in the nonlinear equation, where the coefficients characterize the compression strength. Due to the nonlinearity, this differential equation cannot be solved in an elementary way, unless it is approximated that the column deflection is exceedingly small. Therefore, various methods have been proposed for solving it to realize high-accuracy analysis of columnar buckling behaviors under different mechanical conditions, including the geometry, loading conditions, and boundary conditions \citep{Dinnik1912,Willers1941,FrischFay1961,Gere1962,Ermopoulos1986,Smith1988,Williams1989,Eisenberger1991,Siginer1992,Elishakoff1999,Elishakoff2000,Elishakoff2001,QSLi2001,Duan2008,QSLi2009,Darbandi2010,CYWang2010,DJWei2010,LZhang2016,Akbas2017,BJXiao2019,Batista2019,PZhou2019,HBLy2019,ZPHu2019,Teter2020,Szychowski2020,Chinnawut2020}. Most methods have been based on either the Bessel function, elliptic integral, or other nonelementary functions/integrals, ensuring high accuracy in the analysis of column buckling behavior even with large deflections; those based on generalized hypergeometric functions \cite{Duan2008} and the singular perturbation method \cite{Darbandi2010} are only a few to mention. While such the sophisticated techniques are of fundamental importance, the mathematical formulas derived from them are complex and often beyond intuitive understanding. Using a simpler more manageable formula at the expense of a certain numerical accuracy helps in obtaining an overall picture of the buckling behavior, particularly from an application perspective.

In this study, simple analytical relationships between the concentrated load acting on the tip of the column, the column's own weight, and the column deflection angle are formulated. The estimated formulas concisely reflect the contribution of the tip-concentrated load and self-weight to the tilt angle at the column tip, respectively. In addition, it is numerically observed that the tip-tilt angle is almost a square root of the load increment within the postbuckling state. All the results are within the realm of elementary calculus, providing an intuitive perspective of columnar buckling under combined loading.

\section{Method}

\subsection{Governing equation}

\begin{figure}[ttt]
\centering
\includegraphics[width=0.45\textwidth]{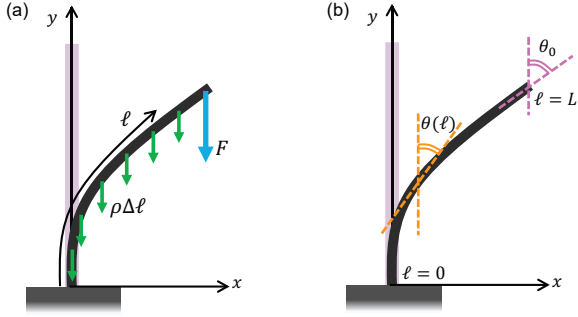}
\caption{Configuration diagram of a column under combined loads: (a) Concentrated load $F$ at the tip and uniformly distributed load $\rho$ caused by the self-weight per length. (b)Tilt angle $\theta(\ell)$ and tip angle $\theta_0$ at $\ell=L$.}
\label{fig01_column_model}
\end{figure}

We consider a vertically built heavy elastic column with a free end at the tip. 
Figure \ref{fig01_column_model}(a) shows the configuration of a column subjected to downward concentrated load at the tip as well as uniformly distributed weight. 
Under the combined load condition, the local bending moment, $M(\ell)$, satisfies
the relation of
\begin{equation}
\frac{dM}{d\ell} = \left\{ F + \rho (L-\ell) \right\} \sin\tht(\ell).
\label{eq_002}
\end{equation}
Here, $F$ is the concentrated downward load applied at the tip, $\rho$ is the self-weight of the column per axial length, and $L$ is the column length; $\ell$ is the arc length measured from the bottom end, and $\theta$ is the local tilt angle. Angle $\theta$ is defined such that $\theta=0$ in the vertical direction and $\theta=\pi/2$ in the horizontal; see Fig.~\ref{fig01_column_model}(b).

The column is sufficiently slender such that the local curvature is proportional to the local bending moment $M(\ell)$ as
\begin{equation}
EI \frac{d \tht(\ell)}{d \ell} = -M(\ell),
\label{eq_001x}
\end{equation}
where $E$ is the Young's modulus
and $I$ is the second moment of area of the column.
For convenience, we introduce three dimensionless parameters \citep{FrischFay1961}:
\begin{equation}
\aaa = \frac{F L^2}{EI}, \quad
\bbb = \frac{\rho L^3}{EI},\quad
s = 1-\frac{\ell}{L},
\end{equation}
where $\aaa$ and $\bbb$ represent the relative importance of the concentrated load $F$
and self-weight $\rho L$, respectively,
to the flexural rigidity $EI$.
The normalized arc length $s$ is defined such that
$s=0$ at the tip and $s=1$ at the bottom.
Using these three dimensionless parameters,
the governing equation is derived from Eqs.~(\ref{eq_002}) and (\ref{eq_001x}) as 
\begin{equation}
\frac{d^2 \tht(s)}{d s^2} = -(\aaa+\bbb s)\sin\tht(s),
\label{eq_001}
\end{equation}
and the associated boundary conditions are 
\begin{eqnarray}
& & \tht(s)=0 \;\;{\rm at}\;\; s=1 \;\; \mbox{({\it i.e.,} at the bottom)}, 
\label{eq_003} \\
& & \frac{d\tht(s)}{ds} = 0 \;\;{\rm at}\;\; s=0 \;\; \mbox{({\it i.e.,} at the tip)}.
\end{eqnarray}

\subsection{Approximation solution based on series expansion}

Our immediate task is to derive an approximate solution for Eq.~(\ref{eq_001}) with respect to $\tht(s)$, depending on $\alpha$ and $\beta$. To realize this, we expand $\theta(s)$ using a Maclaurin series up to the $n$-th order as follows:
\begin{equation}
\tht(s) 
= \tht_0 + s \tht_0' + \frac{s^2}{2!} \tht_0'' + \frac{s^3}{3!} \tht_0^{(3)} + \cdots
+ \frac{s^n}{n!} \tht_0^{(n)},
\label{eq_008}
\end{equation}
where $\tht_0^{(k)}$ indicates that $d^k \tht(s)/d s^k$ at $s=0$.
It is shown below that all the derivatives $\tht_0^{(k)}$ on the right of Eq.~(\ref{eq_008}) can be expressed using functions of $\alpha$, $\beta$, and $\theta_0$.
First, the boundary condition at $s=0$ is 
\begin{equation}
\tht_0'=0.
\end{equation}
Next, from Eq.~(\ref{eq_001}), 
\begin{equation}
\tht_0'' = -\alpha \sin \tht_0.
\end{equation}
For the higher-order derivatives, we use Eq.~(\ref{eq_001}) to reduce the order of differentiation;
for instance,
\begin{eqnarray}
\frac{d^3 \tht(s)}{d s^3}
&=& \frac{d}{ds} \left\{ -(\aaa+\bbb s)\sin\tht(s) \right\} \no \\
&=& -\bbb \sin\tht(s) - (\aaa+\bbb s) \cos\tht(s) \frac{d \tht(s)}{ds},
\end{eqnarray}
implying that
\begin{equation}
\tht_0^{(3)} = -\bbb \sin\tht_0.
\end{equation}
Repeating this procedure up to the ninth order, we obtain
\begin{eqnarray}
\tht_0^{(4)} &=& \frac{\aaa^2}{2} \sin 2\tht_0, \\
\tht_0^{(5)} &=& 2\aaa\bbb \sin 2\tht_0, \\
\tht_0^{(6)} &=& 2\bbb^2 \sin 2\tht_0
+\aaa^3
\left( 
2\sin\tht_0 - \sin 3\tht_0
\right), \\
\tht_0^{(7)} &=& \frac{\aaa^2 \bbb}{2}
\left( 
33 \sin\tht_0 - 17 \sin 3\tht_0
\right), \\
\tht_0^{(8)} &=& \frac{\aaa}{2}\sin\tht_0
\left( 
42 \bbb^2 -15 \aaa^3 \cos\tht_0 \right. \no \\
& &
\left. \quad -98\bbb^2 \cos2\tht_0 +17\aaa^3 \cos3\tht_0
\right), \\
\tht_0^{(9)} &=& \bbb \sin\tht_0
\left( 
21 \bbb^2 - 108 \aaa^3 \cos\tht_0 \right. \no \\
& & \quad \left. - 49 \bbb^2 \cos 2\tht_0 +124 \aaa^3 \cos 3\tht_0
\right).
\end{eqnarray}
In the following discussion, the series expansion is truncated at the ninth degree for securing the numerical accuracy in engineering application. This is because, when $\alpha$ is sufficiently smaller than the unit, the terms of $\theta_0^{(7)}$ and $\theta_0^{(8)}$ hardly contribute to $\theta(s)$ as they have a common factor of $\alpha$. Similarly, when $\beta$ is nearly equal to zero, the terms of $\theta_0^{(5)}$, $\theta_0^{(6)}$, $\theta_0^{(7)}$, as well as $\theta_0^{(9)}$ give only minor contributions to $\theta(s)$ as they have a common factor of $\beta$. Therefore, the series expansion up to the ninth degree is necessary to include the effects of higher-order terms even if either $\alpha$ or $\beta$ is much smaller than the unit.

It is to be noted that the boundary condition at $s=1$ requires $\tht_0^{(k)}$ to satisfy 
\begin{equation}
\sum_{k=0}^n \frac{\tht_0^{(k)}}{k!} = 0,
\label{eq_030}
\end{equation}
as proved by imposing Eq.~(\ref{eq_003}) into Eq.~(\ref{eq_008}).
All $\tht_0^{(k)}$ are functions of $\alpha$, $\beta$, and $\tht_0$ as demonstrated above; hence, Eq.~(\ref{eq_030}) relates the tip-tilt angle $\theta_0$ to the loading conditions determined by an $\alpha$ and $\beta$ pair.

Equation (\ref{eq_030}) may have several roots of $\tht_0$ depending on the values of $\alpha$ and $\beta$. Among them, the least non-negative root is the physically relevant tip-tilt angle of the column. Through actual calculation, we solved Eq.~(\ref{eq_030}) using the bisection method to find the least root. After obtaining $\theta_0$, $\theta(s)$ can be calculated using Eq.~(\ref{eq_008}) for any value of $s$; thereby, the deflection curve of the column can be drawn for a given $\alpha$ and $\beta$.
If the origin of the $x$-$y$ coordinate plane is set at the bottom end of the column,
the deflection curve in the $x$-$y$ plane
can be parametrically expressed as
\begin{eqnarray}
x(s) &=& L \int_s^1 \sin \theta(s^*) ds^*, \\
y(s) &=& L \int_s^1 \cos \theta(s^*) ds^*.
\label{eq_074}
\end{eqnarray}
The integration was numerically performed according to Simpson's $3/8$ rule.

\begin{figure}[ttt]
\centering
\includegraphics[width=0.4\textwidth]{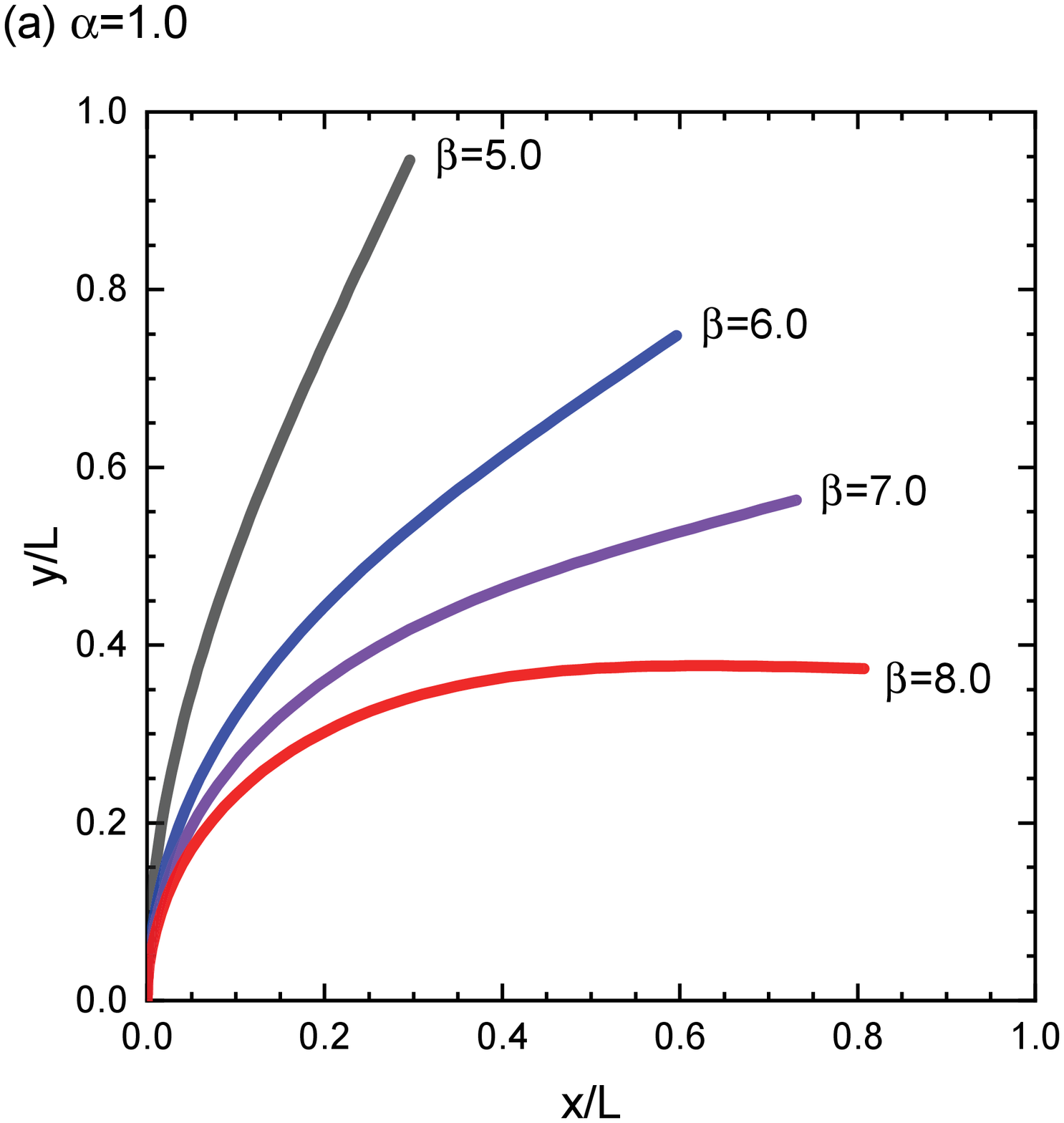}
\includegraphics[width=0.4\textwidth]{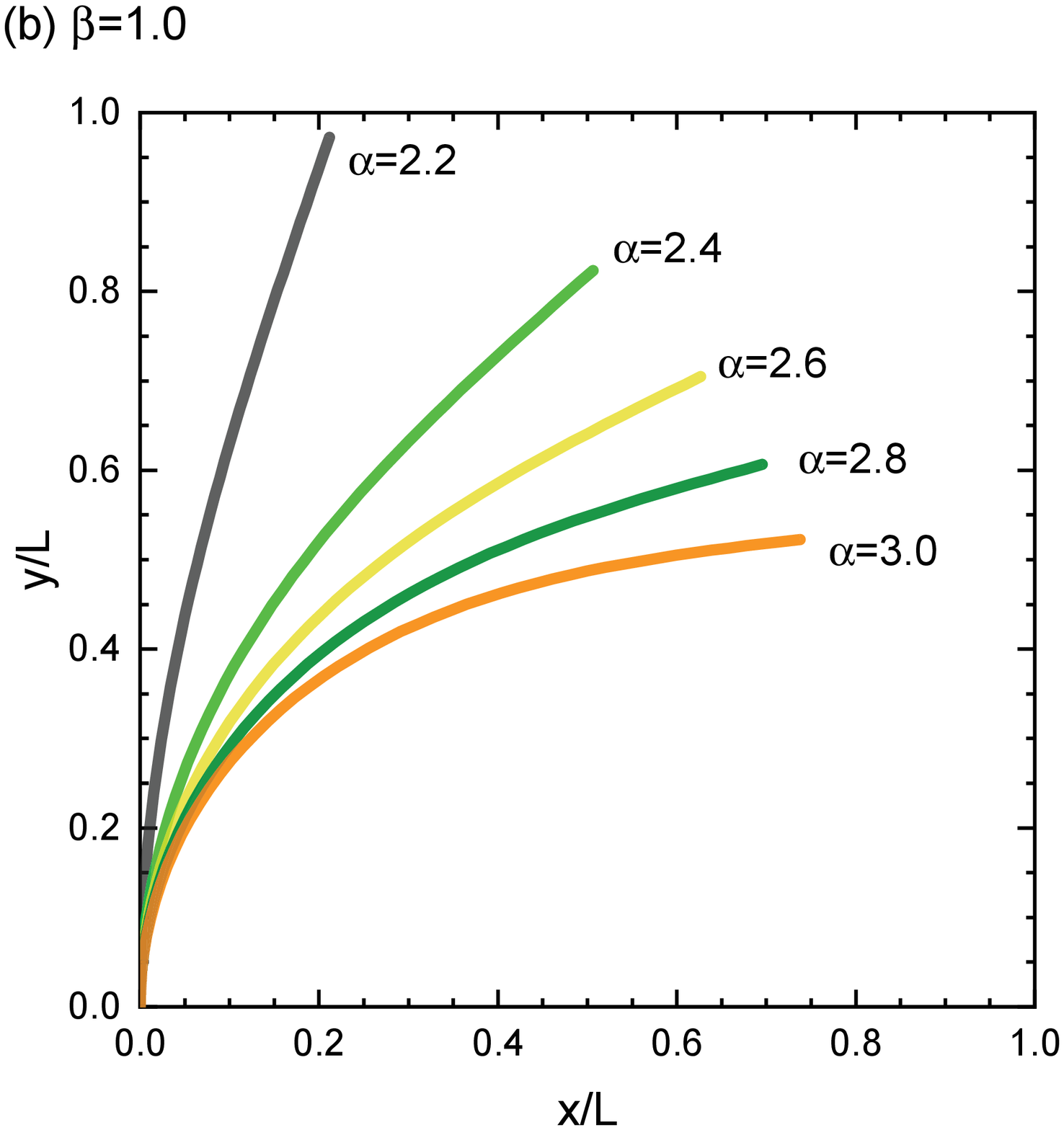}
\caption{Deflection curves for various $\alpha$ and $\beta$ parameter conditions: 
(a)	$\alpha=1.0$ (fixed) and (b) $\beta=1.0$ (fixed).}
\label{fig02_deflection_aorb_fixed}
\end{figure}

\begin{figure}[ttt]
\centering
\includegraphics[width=0.48\textwidth]{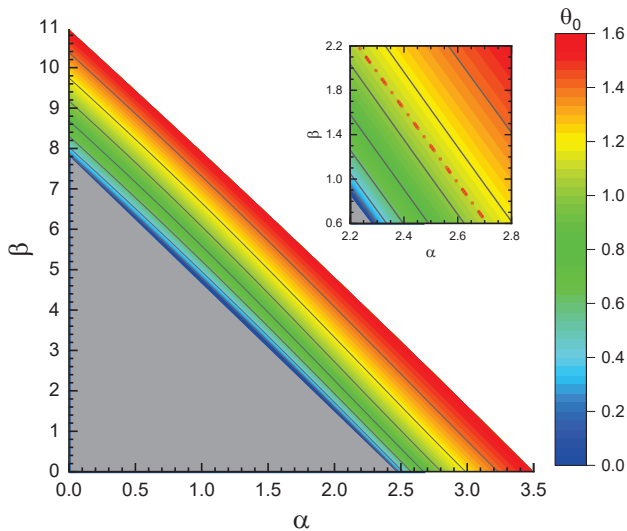}
\caption{Contour map of the tip-tilt angle $\theta_0$ (unit: radian).
Inset: Magnified view of the right-bottom region. See text for the dashed-dotted line.}
\label{fig03_contour_thetazero}
\end{figure}

\section{Result}

\subsection{Relationship between the combined loads and tip-tilt angle}

Figure \ref{fig02_deflection_aorb_fixed} depicts the typical deflection behaviors of the column for various $\alpha$ and $\beta$. When $\alpha$ is fixed at $\alpha=1.0$, as presented in Fig.~\ref{fig02_deflection_aorb_fixed}(a), the column remains vertical until $\beta$ reaches a critical value of $\beta_c = 4.75$. When $\beta$ exceeds $\beta_c$, the column begins to deflect and the tip-tilt angle $\theta_0$ increases monotonically with $\beta$. Note that the threshold value $\beta_c$ changes, if $\alpha$ is fixed at a different value, as demonstrated later. A similar deflection process is observed when $\beta$ is fixed; see Fig.~\ref{fig02_deflection_aorb_fixed}(b). For instance, when $\beta$ is fixed at $\beta=1.0$, the critical value of $\alpha$ is $\alpha_c = 2.16$ beyond which $\theta_0$ increases with $\alpha$. Again, the value of $\alpha_c$ depends on the choice of $\beta$.

\begin{figure*}[ttt]
\centering
\includegraphics[width=0.99\textwidth]{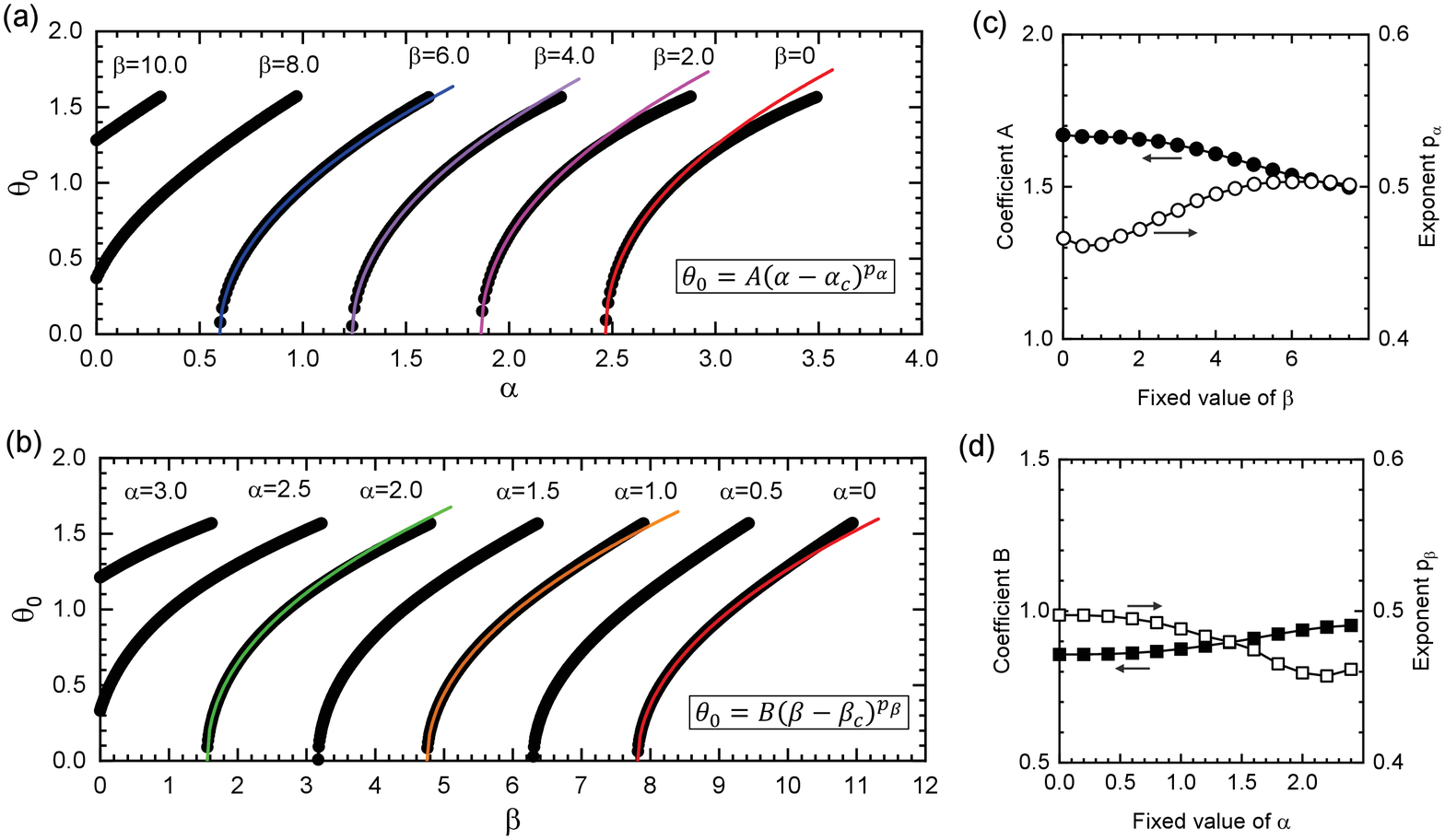}
\caption{(a,b) Power-law dependences of $\theta_0$ on $\alpha$ and $\beta$.
The fitting curves for data points in the $0<\tht_0<1.0$ range are also shown.
(c,d) Optimal values of coefficients $A$ and $B$, and those of exponents
$p_\alpha$ and $p_\beta$.}
\label{fig04_square_behav}
\end{figure*}

The contour map in Fig.~\ref{fig03_contour_thetazero} illustrates the variation of the tip-tilt angle $\tht_0$ with the changes in the values of $\alpha$ and $\beta$. The upper limit of the contour line is set to 1.6 (unit: radian); data above this limit are omitted in the plot. It can be clearly observed that $\theta_0 = 0$ within the left-bottom triangular region (gray colored). This indicates that for relatively small $\alpha$ and $\beta$, the vertically standing state is energetically favored; hence, the column is not deflected but merely compressed in the axial direction. In the rainbow-colored region, in contrast, $\tht_0>0$, implying that deflection will occur. The boundary curve that separates the two equilibrium states $\tht_0=0$ and $\tht_0>0$ is found to be a nearly linear slanted line in the $\alpha$-$\beta$ plane, which has intersections with the $\alpha$-axis and $\beta$-axis, respectively, at $\beta_c=7.815$ and $\alpha_c=2.468$. The latter value is close to the normalized Euler buckling load $\pi^2/4$ with deviation of 0.024\%. If the series expansion in Eq.~(\ref{eq_008}) were truncated at the seventh or fifth degree, for example, the deviation from teh Euler value would be enlarged to 0.097\% or 2.78\%, respectively. It is also noted that our values of $\alpha_c$ and $\beta_c$ obtained by the ninth-degree expansion are nearly equal to those evaluated in the existing work with differences less than 1\% \cite{Duan2008, Darbandi2010}.

An important observation in Fig.~\ref{fig03_contour_thetazero} is that 
all the contour lines are nearly parallel to each other,
approximately obeying the following relationship:
\begin{equation}
\alpha + \frac{10}{3} \beta = c,
\label{eq_016}
\end{equation}
with a $\tht_0$-dependent parameter $c$. The phase boundary, for instance, is nearly equivalent to the line corresponding to $c= 2.47$. Similarly, the dashed-dotted line in the magnified view of the inset of Fig.~\ref{fig03_contour_thetazero} (red colored) coincides with the line for $c= 2.89$. The physical meaning of Eq.~(\ref{eq_016}) is that the effect of the self-weight $\rho L$ on the slender column's deflection is nearly 0.3 times that of the concentrated load $F$ applied at the tip. Hence, as long as the values of $\alpha$ and $\beta$ satisfy Eq.~(\ref{eq_016}), the tip-tilt angle $\tht_0$ will remain constant irrespective of the values of $\alpha$ and $\beta$. This $0.3$-rule represented by Eq.~(\ref{eq_016}), which describes the balance of the contributions by the concentrated load and distributed load to the tip-tilt angle, is the first main finding of this study.

\subsection{Power-law evolution of the tip-tilt angle}

Figure \ref{fig04_square_behav}(a) demonstrates the power-law behavior of $\tht_0$ as a function of $\alpha$ for various fixed values of $\beta$. The series of power-law curves are the edges of the cutting plane that appear when the three-dimensional $\tht_0 (\alpha,\beta)$ surface spanning within the $\aaa$-$\bbb$-$\tht_0$ space (highlighted by the rainbow-colored portion in Fig.~\ref{fig03_contour_thetazero}) is cut along a plane parallel to the $\alpha$-axis (and perpendicular to the page). All the curves are well fitted to the following expression:
\begin{equation}
\tht_0 = A \left( \alpha - \alpha_c \right)^{p_\alpha},
\end{equation}
particularly in the $\tht_0\le 1.0$ range. The threshold $\alpha_c$ decreases monotonically with the increase in $\beta$, as observed in the contour map in Fig.~\ref{fig03_contour_thetazero}. In contrast, the values of the coefficient $A$ and exponent $p_\alpha$ are found to be insensitive to the change in $\beta$; they are approximately estimated as $A\simeq 1.6$ and $p_\alpha \simeq 0.48$, with slight dependence on $\beta$, as shown in Fig.~\ref{fig04_square_behav}(c).

Another class of $\tht_0$ power-law behaviors be deduced when $\alpha$ is fixed and $\beta$ is tuned. Figure \ref{fig04_square_behav}(b) depicts the power-law behavior of $\tht_0$ as a function of $\beta$ for fixed $\alpha$. The fitting curves obey the following expression: 
\begin{equation}
\tht_0 = B \left( \beta - \beta_c \right)^{p_\beta},
\end{equation}
with $B\simeq 0.9$ and $p_\beta\simeq 0.48$. Similar to the previous case, both $B$ and $p_\beta$ are insensitive to the variation in the fixed value of $\alpha$, as shown in Fig.~\ref{fig04_square_behav}(d).

Emphasis should be placed on the fact that the optimal values of $p_\alpha$ and $p_\beta$, both nearly equal to $0.5$, are robust against the variations of $\alpha$ and $\beta$ that characterize the loading conditions. This implies that the tip-tilt angle $\tht_0$ is almost a square root of the increment, if either the concentrated or distributed load is amplified gradually. These nearly square-root behaviors of $\tht_0$ with the load increment can be approximately expressed as
\begin{eqnarray}
& & \tht_0 \propto \sqrt{\alpha - \alpha_c} \;\; \mbox{when $\beta$ is fixed}, \nonumber \\
& & \tht_0 \propto \sqrt{\beta - \beta_c} \;\; \mbox{when $\alpha$ is fixed}.
\label{eq_123}
\end{eqnarray}
Using these approximate formulas, the observed increase in $\tht_0$ with the increase in load can be estimated easily by measuring the initial value of $\tht_0$ before increasing the load. These square root formulas are the second main finding of this study.

\section{Discussion}

We determined that the phase boundary curve
that separates the vertically standing phase ({\it i.e.,} $\theta_0=0$)
and deflection phase ($\theta_0 > 0$) in the $\aaa$-$\bbb$ plane
can be approximately expressed by a simple linear equation (Eq.~(\ref{eq_016})).
To determine the critical values of $\aaa$ and $\bbb$ with higher accuracy, the following formula can be used \citep{CYWang1981}:
\begin{equation}
J_{-\nu}(\zeta_1) J_{-1+\nu}(\zeta_0) + J_{\nu}(\zeta_1) J_{1-\nu} (\zeta_0) = 0
\;\; \mbox{with $\nu=1/3$},
\label{eq_070}
\end{equation}
where $J_{\nu}$ is the $\nu$-th order Bessel function and
\begin{equation}
\zeta_m = \frac23 \left( m+\frac{\alpha}{\beta} \right)^{\frac32} \beta^{\frac12},
\quad m=0,1.
\label{eq_071}
\end{equation}
Using this accurate formula of Eq.~(\ref{eq_070}), we can obtain the $\aaa$-intercept of the boundary curve at $\alpha_c = \pi^2/4$, which is consistent with our approximate value of $\aaa_c=2.46$.
Similarly, the $\bbb$-intercept derived from the accurate formula is $\beta_c = 7.83735$, which is again proximate to our result of $\bbb_c=7.82$. This comparison of the critical values confirms the practicality of our simple expression for the phase boundary, given by Eq.~(\ref{eq_016}).

One of the interesting ways of using the approximate formulas derived in this study is to diagnose the mechanical stability of upright plants, such as large trees and tall bamboos. Understanding the stability of the tree trunk under self-weight and the applied loads caused by the leaf and branches is crucial when developing management strategies to reduce the risk of damage by abiotic agents. In addition, bamboo is a unique plant that does not break, exhibiting considerable flexure, even when subjected to load \citep{Shima2016}; it is extensively used as building material in areas such as Southeast Asia. Note that when using mathematical expressions in actual stability assessment, rapid estimation of the stability using a simple formula is desired, rather than exact numerical data based on expensive numerical simulation. Our simple formula for heavy column deflection facilitates an approximate perspective of the stability analysis of plants in a postbuckling state. However, to investigate the mechanical response of a plant more quantitatively, it is necessary to consider the variation in both the mechanical and geometrical properties of the plant from the ground to the top as well as the effect of initial imperfection. In this context, we need to extend our formulation considering the spatial variation in the cross-sectional geometry \citep{DJWei2010,Shima2020}, that in Young's modulus \citep{PZhou2019}, and the presence of cracks in the initial state \citep{Akbas2017}; I intended to address these interesting problems in future.

The effect of boundary condition variation on the postbuckling behavior is another interesting topic. In this study, our focus was restricted to the deflection of a column in which one end is fixed and the other is free. Another important case to consider is a column in which both ends are fixed because it is frequently used in structural applications, in both macroscopic and nanoscopic \citep{Carr2003,Roodenburg2009,Weick2011} systems. Imposing different boundary conditions on a column can inherently change its mechanical response to the applied loads. Therefore, as in the case of this study, it is interesting to examine whether an approximate expression for the deflection can be derived from a column with fixed conditions at both ends. Our results can pave the way for exploring simple approximate formulas describing column deflection under various end conditions.

\section{Conclusion}

Simple approximate formulas that describe the postbuckling behavior of columns under a tip-concentrated load $F$ and uniformly distributed load $\rho\Delta \ell$ were derived in this study. The main findings include the $0.3$-rule for the balance of the two load contributions to the deflection, given by Eq.~(\ref{eq_016}), and the nearly square root evolution of the deflection angle, given by Eq.~(\ref{eq_123}). These simple formulas relate the degree of deflection to each combined load through a basic function, enabling fast estimation and intuitive understanding of the increase in the deflection angle.

\section*{Acknowledgments}
I would like to thank A.~Takashima for the technical support in manipulating numerical data. Extensive discussions with K.~Ishikawa, S.~Tsugawa, A.~Inoue, Y.~Umeno, and M.~Sato have been illuminating for coming up with the idea of this present study.
This work was supported by JSPS KAKENHI Grant Numbers 
18H03818, 19H02020, 19H05359, and 19K03766.

\bibliographystyle{apsrev4-1}
\bibliography{manu_Shima_arXiv}

\end{document}